\documentclass[preprint,showpacs,preprintnumbers,amsmath,amssymb,nofootinbib]{revtex4}

\usepackage{graphicx}
\usepackage{color}
\usepackage{dcolumn}
\usepackage{bm}

\begin{document}

\preprint{APS/123-QED}

\title{Creatable Universes}

\author{Joan Josep Ferrando}
 \email{joan.ferrando@uv.es}
\author{Ramon Lapiedra}
 \email{ramon.lapiedra@uv.es}
\author{Juan Antonio Morales}%
 \email{antonio.morales@uv.es}
\affiliation{Departament d'Astronomia i Astrof\'{\i}sica,
\\Universitat de Val\`encia, 46100 Burjassot, Val\`encia, Spain.}

\date{\today}

\begin{abstract}
We consider the question of properly defining energy and momenta for
non asymptotic Minkowskian spaces in general relativity. Only spaces
of this type, whose energy, linear 3-momentum, and intrinsic angular
momentum vanish, would be candidates for creatable universes, that
is, for universes which could have arisen from a vacuum quantum
fluctuation. Given a universe, we completely characterize the family
of coordinate systems for which one could sensibly say that this
universe is a creatable universe.
\end{abstract}


\maketitle

\section{Introduction: general considerations}
\label{intro}

Which is the most general universe with null energy, null linear
3-momentum, and null intrinsic 3-angular momentum, and why could
such a question be of interest?

From the early seventies, people have speculated about a Universe
which could have arisen from a quantum vacuum fluctuation
\cite{Albrow}, \cite{Tryon}. If this were the case, one could expect
this Universe to have zero energy.

But, then, why should we consider only the energy? Why not expect
that the linear 3-momentum and angular intrinsic 3-momentum, of a
Universe arising from a vacuum fluctuation, to be zero too? And
finally: why not to expect both, linear 4-momentum and angular
intrinsic 4-momentum, to be zero?

So, in the present paper, we will consider both: linear 4-momentum,
$P^{\alpha} = (P^0,P^i)$, and angular 4-momentum, $J^{\alpha \beta}
= (J^{0i}, J^{ij})$. In all: it could be expected that only those
universes with $P^{\alpha} = 0$, and $J^{\alpha \beta} = 0$, could
have arisen from a quantum vacuum fluctuation. Then, we could say
that only these ones would be `creatable universes'.

Now, as it is well known (see, for example, \cite{Weinberg} or
\cite{Murchadha}), when dealing with an asymptotically flat
space-time, one can define in a unique way its linear 4-momentum,
provided that one uses any coordinate system which goes fast
enough to a Minkowskian coordinate system in the 3-space infinity.

Nevertheless, if, to deal with the Universe as such, we consider
non asymptotically flat space-times, in such space-times these
Minkowskian coordinate systems do not exist. Then, we will not
know in advance which coordinate systems, if any,  should be used,
in order to properly define the linear and angular 4-momentum of
the Universe. This is, of course a major problem, since, as we
will see, and it is well known, $P^{\alpha}$ and $J^{\alpha
\beta}$ are strongly coordinate dependent, and it is so whatever
it be the energy-momentum complex we use (the one of Weinberg
\cite{Weinberg}, or Landau \cite{Landau}, or any other one).

As we have just said, this strong coordinate dependence of
$P^{\alpha}$ and $J^{\alpha \beta}$ is very well known, but, in
spite of this, in practice, it is not always  properly commented or
even taken properly into account. This can be seen by having a look
at the different calculations of the energy of some universes, which
have appeared in the literature (see for example, among other
references, \cite{Johri}, \cite{Banerjee}) since the pioneering
papers by Rosen \cite{Rosen} and Cooperstock \cite {Cooperstock}.

Even Minkowski space can have non null energy if we take non
Minkowskian coordinate systems. This non null energy would reflect
the energy of the \emph{fictitious gravitational field} induced by
such non Minkowskian coordinates, or in other words the energy
tied to the family of the corresponding accelerated observers. So,
in particular, to define the proper energy and momentum of a
universe, we would have to use coordinate systems adapted, in some
sense, to the symmetries of this universe, in order to get rid of
this spurious energy supply. We will address this question in some
detail in the present paper, the summary of which follows.

First, in Sections \ref{sec-2} and \ref{sec-3}, we look for the
family of good coordinate systems in order to properly define the
energy and momenta of the considered universe. Then, given an
arbitrary space-like 3-surface, we uniquely determine the family of
coordinate systems, which are, in principle, good coordinate systems
corresponding to this space-like 3-surface. In Section \ref{sec-4},
under reasonable assumptions, we show that if a given universe has
zero energy and momenta for one coordinate system of the family,
then, it has zero energy and momenta for all coordinate systems of
the family. Furthermore, in Section \ref{sec-5}, under reasonable
assumptions, we show that this ``creatable" character of a given
universe is independent of the above chosen space-like 3-surface. In
Sections \ref{sec-6} and \ref{sec-7} we consider some simple
examples in which we calculate the universe energy and momenta: the
Friedmann-Robertson-Walker (FRW) universes, on one hand, and a
non-tilted Bianchi V universe, on the other hand. Finally, in
Section \ref{sec-8}, we summarize the main results and conclude with
some comments on open perspectives.

Some, but not all, of these results have been presented with
hardly any calculation in the meeting ERE-2006 \cite{ere-06}.

\section{Which coordinate systems?}
\label{sec-2}

We expect any well behaved universe to have well defined energy
and momenta, i. e., $P^{\alpha}$ and $J^{\alpha \beta}$ would be
finite and conserved in time. So, in order for this conservation
to make physical sense, we need to use a {\em physical} and {\em
universal} time. Then, as we have done in \cite{ere-06}, we will
use Gauss coordinates:
\begin{equation}
d s^2 = -dt^2 + dl^2, \quad dl^2 = g_{ij} dx^i dx^j, \quad i,j=
1,2,3.
\end{equation}
In this way, the time coordinate is the proper time and so a {\em
physical} time. Moreover, it is an everywhere synchronized time
(see for example \cite{Landau}) and so a {\em universal} time.

Obviously, we have as many Gauss coordinate systems in the
considered universe (or in part of it) as we have space-like
3-surfaces, $\Sigma_3$. Then, $P^{\alpha}$ and $J^{\alpha \beta}$
will depend on $\Sigma_3$ (as the energy of a physical system in the
Minkowski space-time does, which depends on the chosen $\Sigma_3$,
i.e., on the chosen Minkowskian coordinates).

Now, in order to continue our preliminary inquiry, we must choose
one energy-momentum complex. Since besides linear momentum we will
also consider angular momentum, we will need a symmetric
energy-momentum complex. Then, we will take the Weinberg one
\cite{Weinberg}. This complex has the property that it allows us to
write energy and momenta as some integrals over the boundary
2-surface, $\Sigma_2$, of $\Sigma_3$. Then, any other symmetric
complex with this property, like for example the one from Landau
\cite{Landau}, will enable us to obtain essentially the same results
as the ones we will obtain in the present paper.

Then, taking the above Weinberg complex, one obtains, in Gauss
coordinates, for the linear 4-momentum, $P^{\alpha} = (P^0,P^i)$,
and the angular one, $J^{\alpha \beta} = (J^{0i}, J^{ij})$, the
following expressions \cite{Weinberg}:
\begin{eqnarray}
P^0 & = & \frac{1}{16 \pi G} \int(\partial_j g_{ij} - \partial_i
g)
d \Sigma_{2i},  \label{energy-momentum-a}\\[3mm]
P^i & = & \frac{1}{16 \pi G} \int(\dot{g} \delta_{ij} -
\dot{g}_{ij}) d \Sigma_{2j},
\label{energy-momentum-b}\\[3mm]
J^{jk} & = & \frac{1}{16 \pi G} \int(x_k \dot{g}_{ij} - x_j
\dot{g}_{ki}) d \Sigma_{2i},\label{angular momentum}
\\[3mm]
J^{0i} & = & P^i t - \frac{1}{16 \pi G} \int[(\partial_k g_{kj} -
\partial_j g)x_i + g \delta_{ij} - g_{ij}] d \Sigma_{2j},
\label{angular time momentum}
\end{eqnarray}
where we have used the following notation, $g \equiv
\delta^{ij}g_{ij}$,  $\, \dot{g}_{ij} \equiv
\partial_t g_{ij}$, and where $d \Sigma_{2i}$ is the surface
element of $\Sigma_{2}$. Further, notice, that without losing
generality, the angular momentum has been taken with respect to the
origin of coordinates.

There is an apparent inconsistency in Eqs.
(\ref{energy-momentum-a})-(\ref{angular time momentum}), since we
have upper indices in the left hand and lower ones in the right
side. This comes from the fact that, when deducing these equations
(see Ref. \cite{Weinberg}), starting with the Einstein equations in
its covariant form, $G_{\alpha \beta} = \chi T_{\alpha \beta}$,
indices are raised with the contravariant Minkowski tensor,
$\eta^{\alpha \beta}$. Then, in the right side, one can use
indistinctly upper or lower space indices.

The area of $\Sigma_2$ could be zero, finite or infinite. In the
examples considered next, in Sections \ref{sec-6} and \ref{sec-7},
we will deal with the last two possibilities. In the first case,
when the area is zero, the energy and momenta would be trivially
zero (provided that the metric remains conveniently bounded when
we approach $\Sigma_2$).

\section{More about the good coordinate systems}
\label{sec-3}

From what has been said in the above section, one could erroneously
conclude that, in order to calculate the energy and momenta of a
universe, one needs to write the metric in all $\Sigma_3$, in Gauss
coordinates. Nevertheless, since, according to Eqs.
(\ref{energy-momentum-a})-(\ref{angular time momentum}),
$P^{\alpha}$ and $J^{\alpha \beta}$ can be written as surface
integrals on $\Sigma_2$, all we need is this metric, in Gauss
coordinates, on $\Sigma_2$ and its immediate neighborhood (in this
neighborhood too, since the space derivatives on $\Sigma_2$ of the
metric appear in some of these integrals).

Furthermore, since $P^{\alpha}$ and $J^{\alpha \beta}$ are
supposed to be conserved, we would only need this metric for a
given time, say $t=t_0$. Nevertheless, since in
(\ref{energy-momentum-b})-(\ref{angular time momentum}) the time
derivatives of the metric appear, we actually need this metric in
the elementary vicinity of $\Sigma_3$, whose equation, in the
Gaussian coordinates we are using, is $t=t_0$. Thus, we do not
need our Gauss coordinate system to cover the whole life of the
universe. Nevertheless, in order to be consistent, we will need to
check that the conditions for this conservation are actually
fulfilled (see next the end of Section \ref{sec-4} in relation to
this question).

Now, the surface element $d \Sigma_{2i}$, which appears in the
above expressions of $P^{\alpha}$ and $J^{\alpha \beta}$, is
defined as if our space Gauss coordinates, $(x^{i})$, were
Cartesian coordinates. Thus, it has not any intrinsic meaning in
the event of a change of coordinates in the neighborhood of
$\Sigma_2$. So, what is the correct family of coordinate systems
we must use in this neighborhood to properly define the energy and
momentum of the universe? In order to answer this question, we
will first prove the following result:

\begin{itemize}
\item [] {\it On $\Sigma_2$, in any given time instant $t_0$ there
is a coordinate system such that
\begin{equation} d l_0^2|_{\Sigma_2} = f \delta_{ij} dx^i dx^j \, ,
\quad i,j= 1,2,3,  \label{instantaneous metric on sigma-2}
\end{equation}
where $f$ is a function defined on $\Sigma_2$. That is, the
restriction to $\Sigma_2$ of the 3-metric $d l_0^2 \equiv dl^2
(t=t_0)$ may be expressed in conformally flat form.}
\end{itemize}

The different coordinate systems, in which $d l_0^2|_{\Sigma_2}$
exhibits explicitly its conformal form, are connected to each
other by the conformal group in three dimensions. Then, one or
some of these different conformal coordinate systems are to be
taken as the good coordinate systems to properly define the energy
and momenta of the considered universe. This is a natural
assumption since the conformal coordinate systems allow us to
write explicitly the space metric on $\Sigma_2$ in the most
similar form to the explicit Euclidean space metric. But, which of
all the conformal coordinates should be used? We will not try to
answer this question here in all its generality, since our final
goal in the present paper is to consider universes with zero
energy and momenta. Instead of this, we will give some natural
conditions to make sure that, when the energy and momenta of the
universe are zero in one of the above conformal coordinate
systems, these energy and momenta are zero in any other conformal
coordinate system.

So, according to what we have just stated, we must prove that $d
l_0^2|_{\Sigma_2}$ has a conformally flat form. In order to do
this, let us use Gaussian coordinates, $(y^{i})$ in $\Sigma_3$,
based on $\Sigma_2$. Then, we will have
\begin{equation}
dl_0^2 = (dy^3)^{2} + g_{ab}(y^{3}, y^c) dy^a dy^b \, , \quad
a,b,c= 1,2.
\end{equation}\label{Gauss in sigma3}
In the new $(y^{i})$ coordinates the equation of $\Sigma_2$ is
then $y^{3}=L$, where $L$ is a constant.

Then, taking into account that every 2-dimensional metric is
conformally flat, we can always find a new coordinate system
$(x^{a})$ on $\Sigma_2$, such that we can write $dl_0^2$ on
${\Sigma_2}$, that is to say, $d l_0^2|_{\Sigma_2}$, as:
\begin{equation}
d l_0^2|_{\Sigma_2} = (dy^{3})^2|_{\Sigma_2}+ f(L,x^{a}) \delta_{ab}
dx^a dx^b . \label{Gauss2sigma2}
\end{equation}
Finally, we introduce  the new coordinate
\begin{equation}
x^{3} = \frac{y^{3}-L}{f^{\frac{1}{2}}(L, x^a)} + C,
\label{canvicoord}
\end{equation}
with $C$ an arbitrary constant, which can be seen to allow us to
write $d l_0^2|_{\Sigma_2}$ in the form (\ref{instantaneous metric
on sigma-2}), as we wanted to prove. (Notice that even though, in
the general case, $f$ depends on $x^a$, by differentiating Eq.
(\ref{canvicoord}), one obtains on $\Sigma_2$, that is, for $y^3 =
L$, $d y^3|_{\Sigma_2} = f^{1/2}(L, x^a) d x^3$).

Furthermore, if $r^{2}\equiv\delta_{ij}x^{i}x^{j}$ in the
coordinate system of Eq. (\ref{instantaneous metric on sigma-2}),
and we assume that the equation of $\Sigma_{2}$ in spherical
coordinates is $r=R(\theta,\phi)$, we can expect to have in the
elementary vicinity of $\Sigma_2$:
\begin{equation}
dl^2 = [^0\!g_{ij}(r-R)^{n} + \cdot \cdot \cdot \,  ] \, dx^i
dx^j, \label {2dl-vicinity-sigma2}
\end{equation}
where $n$ is an integer greater than or equal to zero and where
${^0\!g_{ij}}$ are functions which do not depend on $r$.
Furthermore, according to Eq. (\ref{instantaneous metric on
sigma-2}), on $\Sigma_3$, that is, for $t=t_0$, it must be
\begin{equation}
^0\!g_{ij}(r-R)^{n} |_{t=t_0} = f\delta_{ij}.
\label{primer-ordre-sigma2}
\end{equation}

If, leaving aside a boundary at $r=0$, the equation of the
boundary, $\Sigma_2$, is $r=\infty$, we must put $1/r$ where we
have written $r-R$ in the above equation, that is, we will have
instead of (\ref{2dl-vicinity-sigma2}) and
(\ref{primer-ordre-sigma2}):
\begin{equation}
dl^2 = [^0\!g_{ij} r^{-n} + \cdot \cdot \cdot \, ] \, dx^i dx^j,
\quad ^0\!g_{ij}r^{-n} |_{t=t_0} =
f\delta_{ij},\label{primer-ordre-sigma2-r-infi}
\end{equation}
for $r \to \infty$.

The $^0\!g_{ij}$ functions will change when we do a conformal change
of coordinates. But, this is the only change these functions can
undergo. To show this, let us first check which coordinate
transformation, if any, could be allowed, besides the conformal
transformations, if the explicit conformal form of $d
l_0^2|_{\Sigma_2}$ is to be preserved. In an evident notation, these
transformations would have the form
\begin{equation}
x^i = x{^i}'+ y^i(x^j)(t-t_0) \, ,
\end{equation}
in the vicinity of $\Sigma_3$. But it is easy to see that here the
three functions $y^i(x^j)$ must all be zero, if the Gaussian
character of the coordinates has to be preserved. That is, the only
coordinate transformations that can be done on the vicinity of
$\Sigma_2$, preserving on it the metric conformal form
(\ref{instantaneous metric on sigma-2}) and the universal character
of the Gaussian coordinate time, are the coordinate transformations
of the conformal group in the three space dimensions. Thus, we can
state the following result.\footnote{Actually, proving this
uniqueness leads us to consider a family of infinitesimal coordinate
transformations on the vicinity of $\Sigma_2$, which, although
preserving the conformally flat character of the $3$-metric on
$\Sigma_2$, introduce changes in the space derivatives of this
metric on $\Sigma_2$: see the Appendix, at the end of the paper.}

\begin{itemize}
\item [] {\it Given $\Sigma_3$, that is, given the 3-surface which
enables us to build our Gauss coordinates, we have defined uniquely
$P^{\alpha}$ and $J^{\alpha \beta}$, according to Eqs. {\rm
(\ref{energy-momentum-a})-(\ref{angular time momentum})}, modulus a
conformal transformation in the vicinity of $\Sigma_2$.}
\end{itemize}

So, the question is now: how do $P^{\alpha}$ and $J^{\alpha \beta}$
change under such a conformal transformation? As we have said above,
we are not going to try to answer this general question here.
Instead of this, since we are mainly concerned with `creatable
universes', we will explore under what reasonable assumptions the
energy and momenta of a universe are zero for all the above class of
conformal coordinate systems.

\section{Zero energy and momenta irrespective of the conformal coordinates}
\label{sec-4}

The first thing that can easily be noticed concerning the question
is that the global vanishing of $P^{\alpha}$ and $J^{\alpha \beta}$
is invariant under the action of the groups of dilatations and
rotations on $\Sigma_3$.

It is also easy to see that the global vanishing of $P^{\alpha}$ and
$J^{\alpha \beta}$ will be invariant under the translation group on
$\Sigma_3$, provided that one assumes the supplementary condition
$\int \dot{g}_{ij} d \Sigma_{2j}=0$, which is slightly more
restrictive than $P^i = 0$. Actually, this supplementary condition
will be fulfilled in our case, as a consequence of the assumptions
we will make below, in the present section, in order to have
$P^{\alpha}=0$, as we will point out at the end of the section.

In all, we can say that, in the case we are interested here, of
vanishing energy and momenta, $P^{\alpha}$ and $J^{\alpha \beta}$
are invariant under the groups of dilatations, rotations and
translations on $\Sigma_3$.  But all these three groups are
subgroups of the conformal group of coordinate transformations in
three dimensions. Then, we are left with the subgroup of the group
elements that have sometimes been called the \emph{essential}
conformal transformations. But it is known \cite {Krasinski} that
these transformations are equivalent to applying an inversion first,
that is, $r$ going to $1/r$, then a translation, and finally another
inversion. So, in order to see how $P^{\alpha}$ and $J^{\alpha
\beta}$ change when we do a conformal transformation, one only has
to see how they change when we apply an inversion, that is, $r$
going to $r'$, such that
\begin{equation}
r' = \frac{1}{r} \, , \qquad r^2 \equiv \delta_{ij} x^i x^j \, .
\end{equation}
Assume as a first case that the equation of the boundary $\Sigma_2$
is $r=\infty$ plus $r=0$. In this case, the 2-surface element, $d
\Sigma_{2i}$, which appears in the Eqs.
(\ref{energy-momentum-a})-(\ref{angular time momentum}), can be
written as $d \Sigma_{2i}=r^{2} n_{i} d\Omega$, where $n_{i}\equiv
{x^{i}/r}$, and $d\Omega$ is the elementary solid angle.

Now, let us consider the energy first, $P^{0}$. How does it change
when we apply an inversion? This leads us to see how its
integrand,
\begin{equation}\label{integrand-I}
I\equiv r^{2}(\partial_j g_{ij} - \partial_i g) n_{i}d \Omega =
r^{2} (n_{i} \partial_j g_{ij} - \partial_r g) d \Omega,
\end{equation}
changes. After some calculation, one sees that the new value,
$I'$, for $I$ is
\begin{equation}\label{integrand-I-prima}
I'= r^{3}(r\partial_r g - r n_{i} \partial_j g_{ij}+
2n_{i}n_{j}g_{ij}+2g)d\Omega.
\end{equation}
But, the integrands $I$ or $I'$ are both calculated on $\Sigma_2$.
Then, according to Eq. (\ref{primer-ordre-sigma2-r-infi}), $I'$ on
$\Sigma_{2}$ can still be written for $t= t_{0}$ as
\begin{equation}\label{integrand-I-prima-sobre}
I'|_{\Sigma_2}= r^{3}(r\partial_r g - r n_{i} \partial_j g_{ij}+ 8
f)d\Omega.
\end{equation}
In this expression of $I'$ there is a $r^{3}$ common factor. Thus,
if we want $P^{0'}$ to be zero, it suffices that $r^{3}f$ goes to
zero when $r$ goes to $\infty$ and when $r$ goes to zero. In
particular, this means that $f$ must go to zero at least like
$r^{-4}$ when $r$ goes to $\infty$. Then, according to Eq.
(\ref{primer-ordre-sigma2-r-infi}), the functions $g_{ij} - f
\delta_{ij}$, which must go to zero faster than $f$, will go at
least as $r^{-5}$. In a similar way, in order that $r^{3}f$ goes to
zero for $r$ going to zero, $f$ must decrease, or at most cannot
grow faster that $r^{-2}$. In a similar way, $g_{ij} - f
\delta_{ij}$ must decrease for $r$ going to zero, or at most cannot
grow faster than $r^{-1}$. Of course, this asymptotic behavior of
$g_{ij}$ makes the original $P^{0}$ equal zero too. Thus, on the
assumption that the equation of $\Sigma_2$ is $r=\infty$ plus $r=0$,
we have proved that this behavior is a sufficient condition in order
that $P^{0}=0$ be independent of the conformal coordinate system
used.

This natural sufficient condition is not a necessary one, since it
is possible that $P^{0}$ could vanish because of the angular
dependence of $I$. An angular dependence which would make zero the
integral of $I$ on the boundary 2-surface, $\Sigma_2$,
independently of $I$ going to zero or not when $r$ goes to
$\infty$. But, in this case, from (\ref{integrand-I-prima-sobre})
and (\ref{integrand-I}) one sees that the sufficient and necessary
condition to have $P^{0'}$ equal zero is that the integral of $f$
on $\Sigma_2$ be zero because of the special angular dependence of
the function $f$.

Also, one can easily see that, under the above sufficient
conditions, that is, $g_{ij}$ goes to zero at least like $r^{-4}$
for $r \to \infty$, and does not grow faster than $r^{-2}$ for $r
\to 0$, we will have $P^{i}=0$ and $J^{\alpha\beta}=0$,
independently of the conformal coordinate system used. This is so,
because, according to (\ref{primer-ordre-sigma2-r-infi}), this
asymptotic behavior for $g_{ij}$ entails the same asymptotic
behaviour for $\dot{g}_{ij}$.

All in all:

\begin{itemize}
\item[]\it {Under the assumption that the equation of $\Sigma_2$
is $r=\infty$ plus $r=0$, the linear and angular momenta given by
expressions {\rm (\ref{energy-momentum-a})-(\ref{angular time
momentum})} vanish, irrespective of the conformal coordinates used,
if the following sufficient conditions are fulfilled: the metric
$g_{ij}$ of Eq. {\rm (\ref {primer-ordre-sigma2-r-infi})}  goes to
zero at least like $r^{-4}$ for $r \to \infty$ and, on the other
hand, the metric does not grow faster than $r^{-2}$ for $r \to 0$.}
\end{itemize}

In Section \ref{sec-6}, we will see that all this can be applied
to the closed and flat Friedmann-Robertson-Walker (FRW) universes,
whose energy and momenta then become zero.

Let us continue with the question of the nullity of energy and
momenta, leaving now the special case where the equation of
$\Sigma_2$ is $r=\infty$ plus $r=0$ and considering the
complementary case where this equation is $r=R(\theta,\phi)$. Then,
a natural sufficient condition to have energy zero, irrespective of
the conformal system used, is that the exponent $n$ in Eq.
(\ref{2dl-vicinity-sigma2}) be greater or equal to $n=2$. This is a
sufficient condition similar to the one which was present, in a
natural way, in the above case, i.e., when the equation of
$\Sigma_2$ was $r=\infty$ plus $r=0$.

But, according to Eq. (\ref{2dl-vicinity-sigma2}), the above
asymptotic behavior, $n \geq 2$, extends to $\dot{g}_{ij}$. Then,
it can easily be seen that this entails not only the vanishing of
the energy of the considered universe, but also the vanishing of
its linear 3-momentum and angular 4-momentum irrespective of the
conformal coordinate system used.

All in all, we have established the following result:
\begin{itemize}
\item [] {\it Under the assumption that the equation of $\Sigma_2$
is $r=R(\theta,\phi)$, the linear and angular momenta given by
expressions {\rm (\ref{energy-momentum-a})-(\ref{angular time
momentum})} vanish, irrespective of the conformal coordinates used,
if the following sufficient condition is fulfilled: the metric
$g_{ij}$ of Eqs. {\rm (\ref{2dl-vicinity-sigma2})} and {\rm
(\ref{primer-ordre-sigma2})} vanishes fast enough in the vicinity of
$\Sigma_2$. More precisely, the exponent $n$ in Eq. {\rm
(\ref{2dl-vicinity-sigma2})} is greater than or equal to $n=2$.}
\end{itemize}

In some particular cases, a more detailed analysis, than the one we
have just displayed, enables not only sufficient conditions to be
given, but also necessary and sufficient ones, to have zero energy
and momenta irrespective of the conformal coordinate system used.
But we are not going to give these details here since, in any case,
the point will always be to write the space metric, $g_{ij}$, in the
elementary vicinity of $\Sigma_2$ and $\Sigma_3$, in the form of
Eqs. (\ref{2dl-vicinity-sigma2}) and (\ref{primer-ordre-sigma2}) or,
alternatively, in the form of Eq.
(\ref{primer-ordre-sigma2-r-infi}). Once one has reached this point,
one could readily say if, irrespective of the conformal coordinate
system used, the energy and momenta of the universe vanish or not.

Finally, we must realize that, from the beginning of Section
\ref{sec-3}, all what we have said about the proper definition of
energy and momenta of a given universe lies on the basic
assumption that these are conserved quantities. Then, it can
easily be seen that a sufficient condition for this conservation
is that the second time-time and time-space derivatives of the
space metric $g_{ij}$ vanish on $\Sigma_2$ for the generic
constant value, $t_0$ of $t$. But this is entailed by the
asymptotic behavior of $g_{ij}$ assumed in Eq.
(\ref{2dl-vicinity-sigma2}) or Eq.
(\ref{primer-ordre-sigma2-r-infi}). This is the answer to the
consistency question raised at the end of the second paragraph, at
the beginning of Section \ref{sec-3}.

To end the section, notice that the above assumed behavior of $\dot
g_{ij}(t=t_0)$ near $\Sigma_2$ (going like $r^{-4}$, or like $(r -
R)^2$, or even at most like $r^{-2}$ for $r \to 0$, according to the
different cases we have considered) makes not only $P^i =0$, but
also $\int \dot g_{ij} d \Sigma_{ij} = 0$, as we have announced at
the beginning of the section.

\section{The nullity of energy and linear momentum against
a change of $\Sigma_3$} \label{sec-5}

Let us look back at Section \ref{sec-2}, where we have selected a
space-like 3-surface, $\Sigma_3$, from which to build a coordinate
Gauss system. The energy and momenta of the considered universe
are then in relation to the selected 3-surface, that is, depend on
this selected 3-surface. This is not a drawback in itself, since,
as we put forward in that section, the energy of a given physical
system in the Minkowski space also depends on the Minkowskian
observer, and so it depends on the space-like 3-surface associated
to the coordinate system used through the equation $t=t_0$.
Nevertheless, when this energy and the corresponding linear
3-momentum are both zero for a Minkowskian system, then they are
obviously zero for any other Minkowskian system.

Thus, if the definition of null energy-momentum for a given
universe that we have given in the last section is correct, one
could expect that $P^{\alpha}=0$ should remain valid irrespective
of the 3-surface $\Sigma_3$ used.

We will prove this, first in the case where the equation of
$\Sigma_2$ is $r=\infty$ plus $r=0$, and then in the complementary
case where the equation of $\Sigma_2$ is $r=R(\theta,\phi)$.

In the first case, we will assume that the space metric $g_{ij}$
goes to zero at least like $r^{-3}$ when $r \to \infty$ and that it
also behaves {\em conveniently} for $r=0$. Here, ``conveniently"
means that the metric decreases, or at most grows no faster than
$r^{-1}$, when $r$ goes to zero. We can take these assumptions for
granted since in Section \ref{sec-4}, in order to have $P^{0}=0$
irrespective of the conformal coordinate system used, we had to
assume, as a sufficient condition, the behavior $r^{-4}$ for $r \to
\infty$, besides the above convenient behavior for $r=0$. Notice
that the above $r^{-3}$ asymptotic behavior, as any other faster
decaying, when completed with that convenient behavior for $r=0$,
allows us to have $P^{\alpha}=0$. Indeed, with these assumptions, in
Eq. (\ref{energy-momentum-a}), the integrand of $P^{0}$, for $r$
going to $\infty$, will go like $r^{-4}$, and the one of $P^{i}$
like $r^{-3}$. This sort of decaying, plus the above convenient
behavior for $r=0$, will make $P^{0}$ and $P^{i}$ vanish.

Now, imagine that we slightly change $\Sigma_3$, from the original
$\Sigma_3$ to a new $\widetilde{\Sigma}_3=\Sigma_3 +
\delta\Sigma_3$. Then, we will have the corresponding elementary
coordinate change between any two Gauss systems associated to
$\Sigma_3$ and to $\widetilde{\Sigma}_3$, respectively:
\begin{equation}
x^{\alpha}=x^{\alpha'}+\epsilon^{\alpha}(x^{\beta}),
\label{canvi-infinitessimal}
\end{equation}
where $\vert\epsilon^{\alpha}\vert<<\vert x^{\alpha}\vert$, and
where the absolute values of all partial derivatives of
$\epsilon^{\alpha}$ are order $\vert \epsilon\vert<<1$.

Taking into account that $g_{00}=-1$ and $g_{0i}=0$, we will find
for the transformed 3-space metric, to first order in $\epsilon$:
\begin{equation}
g'_{ij}=g_{ij}+g_{ik}\partial_{j} \epsilon^{k} + g_{jk}\partial_{i}
\epsilon^{k}. \label{metrica variada}
\end{equation}
Now, to calculate the new energy, $\widetilde{P}^{0}$,
corresponding to this transformed metric, we will need
$g'_{ij}(t'=t_0)$ in the vicinity of $\widetilde{\Sigma}_2$ (the
boundary of $\widetilde{\Sigma}_3$). According to Eq.
(\ref{metrica variada}), we will have to first order
\begin{equation}
g'_{ij}(t'=t_0)=(g_{ij}+\epsilon^{0}\dot g_{ij}+g_{ik}\partial_{j}
\epsilon^{k} + g_{jk}\partial_{i} \epsilon^{k})(t=t_0),\label{new
metric}
\end{equation}
for any value of $t_0$ and everywhere on $\Sigma_3$.

From this equation we see that $g'_{ij}(t'=t_0)$ goes to zero as
least like $r^{-3}$, when we approach $\Sigma_2$ through $r$ going
to $\infty$, provided that, as we have assumed, $g_{ij}(t=t_0)$ goes
this way to zero. Similarly, for $r \to 0$, $g'_{ij}(t'=t_0)$ will
decrease, or at most will grow no faster than $r^{-1}$, provided we
have assumed that decreasing or this growing respectively, for
$g_{ij}(t=t_0)$.

Furthermore, one can be easily convinced that $g'_{ij}(t'=t_0)$
will keep the same asymptotic behavior when we approach
$\widetilde{\Sigma}_2$ instead of $\Sigma_2$. Indeed, in the
ancient space coordinates, $x^{i}$, previous to the infinitesimal
coordinate change (\ref{canvi-infinitessimal}), the equation of
$\widetilde{\Sigma}_2$ is still $r=\infty$, or more precisely
$r=\infty$ plus $t'=t_{0}$, whereas the equation of $\Sigma_2$ was
$r=\infty$ plus $t=t_0$. (The same can be established for the
other boundary sheet, $r=0$. See, next, the case where the
equation of $\Sigma_2$ is $r=R(\theta, \phi)).$

Then, as we have said, $g'_{ij}(t'=t_0)$ goes to zero as least like
$r^{-3}$, when we approach $\widetilde{\Sigma}_2$ through $r$ going
to $\infty$. This means that the new energy, $\widetilde{P}^0$,
corresponding to the new Gauss 3-surface, $\widetilde{\Sigma}_3$, is
zero, as the original energy was.

On the other hand, because of (\ref{primer-ordre-sigma2-r-infi}),
$\dot g_{ij}(t=t_0)$, as $g_{ij}(t=t_0)$, will go to zero like
$r^{-3}$ when $r \to \infty$, and will decrease, or at most will
grow no faster than $r^{-1}$, when $r \to 0$. Then, also
$\widetilde{P}^i$, and so the entire 4-momentum,
$\widetilde{P}^\alpha$, corresponding to the new 3-surface,
$\widetilde{\Sigma}_3$, is zero, as the original 4-momentum was.

But we can iterate this result along an indefinite succession of
similar infinitesimal shifts of $\Sigma_3$. That is, as we wanted to
prove, $P^\alpha$ will be also zero for the final 3-surface
$\Sigma_3$, which differs now in a finite amount from the original
3-surface. In this way, we could reach any final $\Sigma_3$,
provided that the original and the final metric, in the
corresponding Gauss systems, were regular enough (otherwise we could
not make sure that in all intermediate infinitesimal steps the above
conditions $|\partial_{\alpha} \epsilon^{\beta}|<<1$ could be
satisfied). Here ``regular enough" means that the contribution of
the neighborhood of any metric singularity, which can appear in the
final $\widetilde{\Sigma}_3$, to the calculation of
$\widetilde{P}^{\alpha}$ goes to zero. In this way, we always could
get rid of the difficulty by excluding this neighborhood in the
calculation.

Now, we will prove once more that $P^{\alpha}=0$ is independent of
the chosen 3-surface $\Sigma_3$, this time in the case where the
equation of $\Sigma_2$, the boundary of $\Sigma_3$, is
$r=R(\theta,\phi)$, plus $t=t_0$, instead of $r=\infty$ plus
$t=t_0$. We will prove this under the assumption that the space
metric, $g_{ij}$, goes to zero at least like as $(r-R)^2$ as we
approach $\Sigma_2$. This assumption plays now the role of the above
assumption $g_{ij}$ going like $r^{-3}$ for $r$ going to $\infty$.
Again, in Section \ref{sec-4}, the behavior of $g_{ij}$ and $\dot
g_{ij}$, going like $(r-R)^2$ in the vicinity of $r=R(\theta,\phi)$,
insures that $P^{\alpha}=0$ irrespective of the conformal coordinate
system used. Notice that this assumption makes zero the original
energy-momentum.

Then, as we have done above in the present section, we slightly
change $\Sigma_3$, from this original $\Sigma_3$ to a new space-like
3-surface $\widetilde{\Sigma}_3= \Sigma_{3}+ \delta\Sigma_3$.
Therefore, we will have Eq. (\ref{new metric}). But, this equation
shows that the domain of variation of the space coordinates for the
functions $g'_{ij}$ for $t'=t_0$ is the same that the corresponding
domain for the functions $g_{ij}$ at $t=t_0$. That is, the boundary
of $\widetilde{\Sigma}_3$ is again $r=R(\theta,\phi)$, now for
$t'=t_0$, or, in the ancient coordinate time, for
$t=t_{0}+\epsilon^0$. Of course, to conclude this, we need that the
time derivative of the ancient space metric, $\dot g_{ij}$, be
defined everywhere, that is, be defined all where $g_{ij}$ is
defined. But this must be taken for granted if we assume that the
metric components are functions of class $C^1$ (i. e., its first
derivatives exist and are continuous). This condition holds
independently of the coordinate system used if, as usual,  the
space-time is considered as a differentiable manifold of class $C^2$
(see, for example, Ref. \cite{lich}).

The next step in our proof is to show that $g'_{ij}$ goes also
like $(r-R)^2$, in the vicinity of $\widetilde{\Sigma}_2$. But,
this becomes obvious from Eq. (\ref{new metric}), once one has
proved, as we have just done, that the equation of
$\widetilde{\Sigma}_2$ is $r=R(\theta,\phi)$ plus $t'=t_0$. Thus,
the new energy momentum, $\widetilde{P}^\alpha$, corresponding to
the new 3-surface, $\widetilde{\Sigma}_3$, is also zero.

Finally, to end the proof, we need to check that, for any chain of
consecutive elementary shifts of the original $\Sigma_3$ space-like
surface, leading to a final new $\widetilde{\Sigma}_3$ space-like
surface, we can iterate indefinitely the above procedure of
obtaining, each time, a new energy-momentum which vanishes. But,
this is again obvious from Eq. (\ref{new metric}), since, as we have
assumed, our space-time is a differentiable manifold of class $C^2$,
which entails that for every shift the time derivative of the space
metric, in any admissible coordinate system, is defined wherever the
space metric is defined. Thus, iterating indefinitely the above
procedure, we find that the final energy-momentum, corresponding to
the new space-like 3-surface, $\widetilde{\Sigma}_3$, is also zero,
as we wanted to prove.

Let us specify, all the same, that to reach this conclusion we need
to assume that the metric is ``regular enough". According to what
has been explained above, in the present section, a ``regular
enough" metric is one such that the same metric and its first
derivatives have no singularities, or one such that, in the case
where some of these singularities are present, the contribution of
its neighborhoods to the integrals which define $P^\alpha$ and
$J^{\alpha\beta}$ in (\ref{energy-momentum-a})-(\ref{angular time
momentum}) goes to zero when the areas of these neighborhoods go to
zero.

All in all, under this regularity assumption, we have proved the
following proposition:

\begin{itemize}
\item[]{\it Let it be any two different space-like 3-surfaces,
$\Sigma_3$ and $\widetilde{\Sigma}_3$. Assume that the Gauss metric
$g_{ij}$ built from the original 3-surface, $\Sigma_3$, is ``regular
enough", and that as we approach its boundary $\Sigma_2$ this metric
satisfies:

(i) If the equation of $\Sigma_2$ is $r = \infty$ plus $r=0$,
$g_{ij} \to 0$ at least like $r^{-3}$ when $r \to \infty$ and
$g_{ij}$ decreases, or at most grows no faster than $r^{-1}$, when
$r \to 0$.

(ii) If the equation of $\Sigma_2$ is $r = R(\theta, \phi)$, $g_{ij}
\to 0$ at least like $(r-R)^2$.

Then, the original linear 4-momentum corresponding to the 3-surface
$\Sigma_3$ vanishes, and the linear 4-momentum corresponding to the
other surface, $\widetilde{\Sigma}_3$ vanishes too}.
\end{itemize}

By nearly making the same assumptions and by reproducing the same
reasoning, we have applied in the case of $P^\alpha$, in the new
case of $J^{\alpha \beta}$, one can easily be convinced that, if
$J^{\alpha \beta}$ vanishes for a given 3-surface, $\Sigma_3$, it
will vanish too for any other space-like 3-surface
$\widetilde{\Sigma}_3$. The only change we have to introduce in the
above assumptions, to reach this conclusion, is the following one.
When the equation of $\Sigma_2$ is $r=\infty$, one has to assume
that $g_{ij}(t=t_0)$ goes to zero like $r^{-4}$ instead of $r^{-3}$.
Remember, nevertheless, that this $r^{-4}$ behavior for
$g_{ij}(t=t_0)$ is already what we had assumed in Sec. \ref{sec-4},
in order to have $P^0 = 0$ irrespective of the conformal coordinates
used in $\Sigma_3$.

\section{The example of FRW universes}
\label{sec-6}

As it is well known, in these universes one can use Gauss
coordinates such that the 3-space exhibits explicitly its
everywhere conformal flat character:

\begin{equation}
dl^2 = \frac{a^2(t)}{\left[1+\frac{k}{4}r^2\right]^2} \delta_{ij}
dx^i dx^j \, , \quad r^2 \equiv \delta_{ij} x^i x^j \,
,\label{conformal metric}
\end{equation}
where $a(t)$ is the expansion factor and $k = 0, \pm 1$ is the index
of the 3-space curvature.

Then, this conformally flat character will be valid, \emph{a
fortiori}, on any vicinity of $\Sigma_3$ and $\Sigma_2$. Therefore,
according to Section \ref{sec-3}, we can apply our definitions to
the metric (\ref{conformal metric}). Taking into account Eqs.
(\ref{energy-momentum-a})-(\ref{angular time momentum}), we will
have then:
\begin{eqnarray}
P^0 & = & - \frac{1}{8 \pi G} \int r^2 \partial_r f d \Omega
, \label{conformal energy-momentum-a}\\
P^i & = & \frac{1}{8 \pi G} \int r^2 \dot{f} n_i d \Omega,
\label{conformal energy-momentum-b}\\
J^{jk} & = & \frac{1}{16 \pi G} \int r^2 \dot{f} (x_{k} n_j -
x_{j}
n_k) d \Omega ,\\
J^{0i} & = & P^i t - \frac{1}{8 \pi G} \int r^2 (f n_i - x_{i}
\partial_r f) d \Omega
\end{eqnarray}
with $\, d \Omega = \sin \theta\,  d \theta \, d \phi $, $\, n_i
\equiv x^i/r$, and where we have put
\begin{equation}
f\equiv \frac{a^2(t)}{\left[1+\frac{k}{4}r^2\right]^2}
\end{equation}
which, excluding the limiting case $k=0$, goes as $1/r^4$ for $r \to
\infty$. This is just the kind of behavior that we have assumed in
Section \ref{sec-4} in order to reach the conclusion that $P^\alpha
= 0$, $J^{\alpha\beta}=0$, are conformally invariant. It is also a
behavior which allows to make this vanishing of $P^\alpha$ and
$J^{\alpha \beta}$ independent of the 3-surface, $\Sigma_3$, chosen.

Then, one can easily obtain the following result, in accord with
most literature on the subject (see the pioneering Ref.
\cite{Rosen}, and also Ref. \cite{xulu} for a concise account),
\begin{equation}
\begin{array}{ll}
k = 0, +1: & \quad P^{\alpha} = 0, \quad J^{\alpha\beta} = 0
\end{array}
\end{equation}
that is, the flat and closed FRW universes have vanishing linear and
angular momenta.

Contrary to this, in the case where $k=-1$, one finds for the
energy, $P^{0}=-\infty$. This is because now the metric is singular
for $r= 2$. Thus, in order to calculate its energy, we must consider
the auxiliary universe which results from excluding the elementary
vicinity $r= 2 \pm \epsilon$. Therefore, we will calculate the
energy of this auxiliary universe and then we will take the limit
for $\epsilon \to 0$. But now, the boundary of the 3-space universe
described by this auxiliary metric is double. On the one hand, we
will have, as in the case of $k = 0, +1$, the boundary $r= \infty$,
and on the other hand the new boundary $r=2$ that we can approach
from both sides $r=2 \pm \epsilon$. Both boundaries must be taken
into account when doing the calculation of $P^{0}$ according to the
Eq. (\ref{conformal energy-momentum-a}). Then, it can easily be seen
that the contribution to the energy calculation from the first
boundary, $r= \infty$, vanishes, but further elementary calculation
shows that the contribution from the other boundary is $-\infty$.
Thus, as we have said, the FRW universes with $k=-1$, have
$P^{0}=-\infty$.

All in all, the flat and closed FLRW universes are `creatable
universes', but the open one is not.

\section{The example of some Bianchi universes}
\label{sec-7}

Let us consider the case of the family of non-tilted perfect fluid
Bianchi V universes \cite {bona}, whose metric can be written as
\begin{equation}
ds^{2}=-dt^{2}+A^{2}dx^{2}+ e^{2
 x}(B^{2}dy^{2}+C^{2}dz^{2}),
\end{equation}
where $A$, $B$ and $C$ are functions of $t$.

The first thing one must notice about this universe metric is
that, as in the above case of the FRW universes, it is written in
Gauss coordinates, which according to Section \ref{sec-2} is the
coordinate system family with which to define the proper energy
and momenta of a given universe.

Then, for $t=t_0$, we will have
\begin{equation}
d l_0^2 \equiv dl^2 (t=t_0)=dx^{2}+e^{2\alpha x}(dy^{2}+ dz^{2}),
\end{equation}
where we have rescaled  the original notation $(x,y,z)$ according to
${A_{0}x} \to x$, ${B_{0}y} \to y$, ${C_{0}z} \to z$, and where
$\alpha= 1/A_{0}$, with $A_{0}\equiv {A(t_{0})}$, and so on.
Now, let us move from the variable $x$ to new variable $x'$: $x'=
e^{-\alpha x}/\alpha$. Then, we will have for the instantaneous
space metric, $dl_{0}^{2}$,
\begin{equation}
d l_0^2=\frac {1}{\alpha^{2}x'^{2}}(dx'^{2}+dy^{2}+dz^{2}),
\end{equation}
or changing the above notation such that $x' \to {x}$:
\begin{equation}
dl_0^2=\frac {1}{\alpha^{2}x^{2}}\delta_{ij} dx^{i}dx^{j}.
\end{equation}
This is a conformal flat metric not only in the vicinity of
$\Sigma_2$ but everywhere on $\Sigma_3$ (except for $x=0$). Then,
according to Section \ref{sec-3}, we can use this particular
expression of $dl_{0}^{2}$ to calculate the energy of our family
of Bianchi universes, since, in fact, to calculate this energy we
only need the instantaneous space metric in the vicinity of
$\Sigma_2$.

Now, this metric has a singularity for $x=0$. Thus, in order to
calculate its energy, we must proceed as in the above case of an
open FRW universe. So, we consider the auxiliary universe which
results from excluding the elementary vicinity of $x=0$, $x \in (0,
+\epsilon)$, where we have taken $\alpha > 0$. Therefore, we will
calculate the energy of this auxiliary universe and then we will
take the limit for $\epsilon \to 0$. The boundary of the 3-space
universe described by this auxiliary metric is double. On the one
hand, we will have the boundary $x=+ \infty$, and on the other hand
the boundary $x= + \epsilon$. Both must be taken into account when
doing the calculation of $P^{0}$ according to the Eq.
(\ref{conformal energy-momentum-a}).

Then, it is easy to see that the contribution to $P^{0}$ of the
second boundary, $x= \epsilon$, gives $+\infty$, and that the
contribution of the first boundary, $x=+ \infty$, gives $+\infty$
too. Therefore, we can conclude that the energy of our Bianchi V
family of universes is $P^{0} =+\infty$. Then, this family of
universes, next to the open FRW universe we have just seen, are
examples of non ``creatable universes".

\section{Discussion and prospects}

\label{sec-8} We have analyzed which family of coordinate systems
could be suitable to enable the linear and angular 4-momenta of a
non asymptotically flat universe to be considered as the energy and
momenta of the universe itself, without the spurious energy and
momenta of the fictitious gravitational fields introduced by
accelerated (non inertial) observers. Though we have not been able
to uniquely determine this family in the general case, we have been
able to do so in a particular but interesting case, where the energy
and momenta of the universe vanish. As a consequence, the notion of
a universe having zero energy and momenta is unique and so makes
sense. This result is in contrast with the exhaustive studies on the
energy and momentum of a 3-surface $\Sigma_3$, in General
Relativity, mainly focussed on the asymptotically flat behavior of
$\Sigma_3$ (see \cite{Murchadha} and references therein).

Universes whose energy and momenta vanish are the natural
candidates for universes that could have risen from a vacuum
quantum fluctuation. Here we have called these universes
``creatable universes".

Any given universe could be rejected from the very beginning, as a
good candidate for representing our real Universe, in the event that
it were a non ``creatable" one. We could reject it either before the
inflationary epoch, or after this epoch, or just right now. This
could be the main interest of the characterization of the
``creatable universes" that we have reached in the present paper.
Thus, for example, people have considered the possibility that our
present Universe could be represented by Stephani universes
\cite{Coll,Clarkson,Stelmach,Ferrando}, that is, by a universe which
at different times admits homogenous and isotropic space-like
3-surfaces whose curvature index can change. Such a possibility is a
generalization of the FRW universes and could not be easily
discarded on the grounds of present cosmic observations.
Nevertheless, if all, or some, of these Stephani universes were non
``creatable universes", we could reject them on the grounds of the
assumption that all candidate universes able to represent our real
Universe should be ``creatable universes". This is why it could be
interesting to see which Stephani universes have zero energy and
momenta. For similar reasons, it could be interesting to make the
same analysis in the case of Lema\^{\i}tre-Tolman universes
\cite{Alnes,Chung,Paranjape}, and in the case of a particular
Bianchi type VII universe \cite{Jaffe}. We expect to consider these
questions in detail elsewhere shortly.

\begin{acknowledgments}
This work has been supported by the Spanish Ministerio de
Educaci\'on y Ciencia, MEC-FEDER project FIS2006-06062.
\end{acknowledgments}

\appendix

\section{}
\label{sec-9}
\vspace{-4mm}
{\bf On the uniqueness of the energy and momenta of the Universe
under a coordinate change in the vicinity of the boundary, ${\bf
\Sigma_2}$}
\vspace{3mm}

In Section \ref{sec-3} we claim that the defined energy and momenta
of the Universe, for a given space-like 3-surface, $\Sigma_3$, are
unique, modulus a conformal transformations on the boundary,
$\Sigma_2$, of $\Sigma_3$.

Actually, as we are going to see, proving this uniqueness needs to
consider other coordinate transformations in the vicinity of
$\Sigma_2$ than the ones considered in that section.

Imagine that, according to the protocol we have displayed above to
calculate the proper energy and momenta of the Universe, we have
been able to build the coordinate system in which the $3$-metric,
$g_{ij}$, has a conformally flat form on $\Sigma_2$:
\begin{equation}\label{A1}
g_{ij}|_{\Sigma_2} = f \delta_{ij}
\end{equation}
Let it be $\phi(x^i)=0$ the equation of $\Sigma_2$ in this
coordinate system. Then, change this coordinate system, in the
vicinity of $\Sigma_2$, according to the infinitesimal coordinate
transformation:
\begin{equation}\label{A2}
x^i = x{^i}' + \xi^i(x^j) \phi^2 \equiv x{^i}' + \epsilon^i(x^j),
\end{equation}
where $\xi^i$ are three arbitrary bounded functions in this
vicinity. Notice that in Sec. \ref{sec-3} we only have considered
transformations such as (\ref{A2}) which were linear in $\phi$.
(Infinitesimal transformations such as (\ref{A2}), but with terms
like $\phi^n$, $n>2$, would be irrelevant for our purposes).

The Jacobian matrix of the coordinate transformation (\ref{A2}) is
\begin{equation}\label{A3}
\frac{\partial x^i}{\partial x{^j} '} = \delta^i_j + 2 \, \phi \,
\xi^i \, \partial_j \phi + O(\phi^2).
\end{equation}
Then, we have
\begin{equation}\label{A4}
\frac{\partial x^i}{\partial x{^j} '}|_{\Sigma_2} = \delta^i_j\, ,
\end{equation}
and the coordinate change (\ref{A2}) cannot change the $3$-metric on
$\Sigma_2$, i. e., cannot change Eq. (\ref{A1}): Nevertheless, what
appears in the expression (\ref{energy-momentum-a}) of $P^0$, in Sec
\ref{sec-2}, is not the $3$-metric $g_{ij}$ on $\Sigma_2$, but its
space derivatives, which do change under a coordinate transformation
such as (\ref{A2}). Thus, let us see which this change looks like.
First of all, we will have for the new components $g_{ij}'$ of the
$3$-metric

\begin{equation}\label{A5}
g_{ij}' = g_{ij} + g_{ik} \partial_j \epsilon^k + g_{jk} \partial_i
\epsilon^k = g_{ij} + 2 \, \phi \, \xi^k (g_{ik} \partial_j \phi +
g_{jk} \partial_i \phi) + O(\phi^2).
\end{equation}

On the other hand, from (\ref{A4}) we have
\begin{equation}\label{A6}
\big(\frac{\partial g_{ij}'}{\partial x_k'}\big)_{|_{\Sigma_2}}
\equiv
\partial_k ' g_{ij} '|_{\Sigma_2} = \partial_k g_{ij} '|_{\Sigma_2}.
\end{equation}

Having this in mind, from (\ref{A5}) and (\ref{A1}), we have for
$\partial_k ' g_{ij} '|_{\Sigma_2}$\, :
\begin{equation}\label{A7}
\partial_k ' g_{ij} '|_{\Sigma_2} = \partial_k g_{ij} + 2 f
\partial_k \phi (\xi_i \partial_j \phi + \xi_j \partial_i \phi )\, ,
\end{equation}
where, without confusion,  we have dropped the symbol $|_{\Sigma_2}$
in the right hand side.

Then, according to Eq. (\ref{energy-momentum-a}) in Sec.
\ref{sec-2}, the integrand corresponding to the new energy,
$P^{0'}$, related to the new coordinates $\{x_i'\}$, will be:
\begin{equation}\label{A8}
\partial_j'( g_{ij}' - g' \delta_{ij})|_{\Sigma_2} = \partial_j( g_{ij} - g
\delta_{ij}) + 2f [(\vec{\nabla} \phi)^2 \xi_i - (\vec{\xi} \cdot
\vec{\nabla} \phi) \partial_i \phi] \, ,
\end{equation}
where, again, we have dropped the symbol $|_{\Sigma_2}$ on the right
side.

On the other hand, one has
\begin{equation}\label{A9}
d \Sigma_{2i}' \propto \partial_i' \phi =  \partial_i \phi + O(\phi)
\propto d \Sigma_{2i} + O(\phi) \, .
\end{equation}
According to this and to Eq. (\ref{A8}), we have finally for
$P^{0'}$:
\begin{equation}\label{A10}
P^{0'} = P^0 + \frac{1}{8 \pi G} \int f [(\vec{\nabla} \phi)^2 \xi_i
- (\vec{\xi} \cdot \vec{\nabla} \phi) \partial_i \phi] d \Sigma_{2i}
\, .
\end{equation}
But, since $d \Sigma_{2i}  \propto \partial_i \phi$, we have
\begin{equation}\label{A11}
[(\vec{\nabla} \phi)^2 \xi_i - (\vec{\xi} \cdot \vec{\nabla} \phi)
\partial_i \phi] d \Sigma_{2i} \propto
[(\vec{\nabla} \phi)^2 \xi_i - (\vec{\xi} \cdot \vec{\nabla} \phi)
\partial_i \phi] \partial_i \phi = 0 \, ,
\end{equation}
for all functions $\xi^i$. In all, we have
$$
P^0 = P^{0'}
$$
for any coordinate change such as (\ref{A2}). Then, we have
established, modulus a conformal transformation on $\Sigma_2$, the
uniqueness of $P^0$ for any space-like 3-surface $\Sigma_3$.

In an analogous way, one can complete the proof of the uniqueness of
the components $J^{0i}$ of the angular 4-momentum $J^{\alpha
\beta}$. As long as the uniqueness of $P^i$ and $J^{ij}$ is
concerned, the coordinate transformations (\ref{A2}) leave $P^i$ and
$J^{ij}$ invariant since the space derivatives of the 3-metric
$g_{ij}$ do not appear neither in $P^i$, nor in $J^{ij}$.

\end{document}